\documentclass[letterpaper]{article}
\usepackage{iccc}
\usepackage[numbers]{natbib}
\usepackage{times}
\usepackage{helvet}
\usepackage{courier}
\title{\textbf{Vocal Melody Construction for Persian Lyrics \\
Using LSTM Recurrent Neural Networks}}
\author{Farshad Jafari, Farzad Didehvar, Amin Gheibi\\
Mathematics and Computer Science Department\\
Amirkabir University of Technology\\
Tehran, Iran\\
jafarifarshad@aut.ac.ir, didehvar@aut.ac.ir, amin.gheibi@aut.ac.ir\\
}
\usepackage{epigraph}
\usepackage{graphicx}
\usepackage{float}
\usepackage{multirow}
\usepackage{booktabs}
\usepackage{array}
\usepackage{url}
\newcolumntype{L}{>{\centering\arraybackslash}m{0.5\linewidth}}
\newcolumntype{C}{>{\centering\arraybackslash}m{0.15\linewidth}}
\newcolumntype{D}{>{\centering\arraybackslash}m{0.1\linewidth}}

\usepackage[autostyle=false, style=english]{csquotes}
\MakeOuterQuote{"}

\usepackage{fontenc}

\begin{document}
\maketitle

\begin{abstract}
\noindent The present paper investigated automatic melody construction for Persian lyrics as an input. It was assumed that there is a phonological correlation between the lyric syllables and the melody in a song. A seq2seq neural network was developed to investigate this assumption, trained on parallel syllable and note sequences in Persian songs to suggest a pleasant melody for a new sequence of syllables. More than 100 pieces of Persian music were collected and converted from the printed version to the digital format due to the lack of a dataset on Persian digital music. Finally, 14 new lyrics were given to the model as input, and the suggested melodies were performed and recorded by music experts to evaluate the trained model. The evaluation was conducted using an audio questionnaire, which more than 170 persons answered. According to the answers about the pleasantness of melody, the system outputs scored an average of 3.005 from 5, while the human-made melodies for the same lyrics obtained an average score of 4.078.
\end{abstract}


\section{Introduction}

Over the last few years, artistic production by artificial intelligence had great breakthroughs. These products can help artists in the creation process or even be individual art pieces. Music creation by AI is one of the most regarded fields for researchers due to its sequential nature and recent advancements in neural networks maintaining long-term dependencies.
\par Different subjects can condition music generation. One of the essential conditioners is lyrics because it is one of the most common inspiration sources for human-being composing a song. A song usually consists of words sung by a singer. Singing as an instrument contains a melody, rhythm, and text in a natural language called a lyric. By accepting the linguistic nature of music, the closest exposure of natural language with the music language occurs in singing. The present research assumed that this exposure is not a coincidence, and expressing these two languages parallel leads to their correlation.


\subsection{General Objectives}

The present study investigated the possibility of composing a melody for a given text using a neural network trained on a parallel corpus of lyric-notes. In other words, a software that takes a lyric in Persian as the input and suggests a melody as the output. 
Our motivation to do this research is mainly pursuing these three objectives:

\begin{itemize}
\item \textbf{Iranian music dataset}: Due to the lack of a digital Persian music dataset, the present study seeks to provide an efficient and accurate collection.
\item \textbf{Implementing an automatic melody composer for Persian lyrics}: Classical Persian music does not follow the Western way of choosing scales and chords. Besides, quarter-tones are commonly used in Persian music, which does not even happen in Western music. Furthermore, the tremolo technique is an inseparable part of vocals in Persian music. The solution we propose could be one of the premiers in this field for Persian music. The system will use open-source deep learning software modified to suit the scope of the problem.
\item \textbf{Evaluating the proposed solution}: Automatic evaluation is done using machine translation evaluation methods. The system also needs to be evaluated by humans, which is not as straightforward as automatic evaluation. For instance, human evaluation of digital outputs is impossible, so they must be performed and recorded as audible pieces.
\end{itemize}

\section{Related Works}

In recent years, several solutions for digital music composition have been developed. Generally, every computational solution for sequential problems can be used for music due to its sequential nature. At first, statistical solutions such as Random forest and Markov chain were used. However, with advancements in neural networks, the previous statistical solutions are hardly used. There are many systems developed regarding music composition; however, the present research investigated Persian music composition for Persian lyrics due to a lack of sufficient studies in this field. Fukayama has proposed a probabilistic model for melody generation, observing the pitch alterations in the prosody of Japanese words \cite{fukayama10}. Ackerman suggests a system for generating melody and rhythm based on Random forests \cite{ackerman17}. Using neural solutions to solve this problem is not yet very common. One of the few articles approaching this method is \cite{bao19} which uses a hierarchical RNN melody decoder to generate proper notes for a given lyric.
All the methods we have named are music generation based on musical symbols. There are different methods for solving this problem based on raw audio files. For example, the paper \cite{dhariwal20} has trained a model that generates impressive pieces by processing audio files and compressing them using HQ-VAE (Quantised-Variational AutoEncoder). Dhariwal et al. have managed to condition the machine's output based on lyrics, genre, and artist. They had the audio files and the lyrics of the songs separately, and one of the challenges was to align the lyrics on the song.

\section{Methodology}
This research approaches the problem mainly by machine translation methods. In other words, lyric and music are considered two languages with translation capabilities. Bao modeled a three-layer encoder of the source language (musical notes) in which each layer holds the pitch, duration, and rhythm, respectively. In this paper we use only one encoder layer. For holding the information about pitch and rhythm, the smallest unit in music language is considered as "pitch + octave + duration," so a one-layer encoder in our architecture will do the same in storing information about musical features. In Bao's article, keeping alignment information led the note sequence to be dependent on lyrics alignment; in other words, the source and destination grammar were assumed to be the same. However, this assumption might not be correct.
By removing the alignment information, it was assumed that a sentence in the destination language is not necessarily obtained by translating word by word from the source language, so the translation of a word can be in any other position in the destination language. For instance, in a sentence of a lyric, the Persian word "Ay" appears, singing the high note in the melody is expected. Because semantically, "Ay" evokes pain and screaming, usually sung in a higher pitch.
Despite the lack of alignment information in training data, we can find proper alignments for output notes based on lyrics by having an attention mechanism in the model. As one of the crucial achievements of Bao's article, this point leads to the realization of the one-to-many relationship between syllables and notes.

\par The general steps of the melody generation system developed for this paper are as follows:
\begin{enumerate}
\item Automatic recognition of musical notes from the book note images.
\item Human editing to correct the mistakes of the previous step and integrate the phonetics of Persian letters in the lyrics.
\item Saving the output of the previous step in MusicXML format.
\item Extracting the selected features for the training process from note sheets in MusicXML.
\item Creating a parallel corpus of lyrics and notes with proper segmentation of melodic sentences.
\item Training a sequence-to-sequence neural network using LSTM cells.
\end{enumerate}
\begin{figure}[H]
\centering
\includegraphics[width=.45\textwidth]{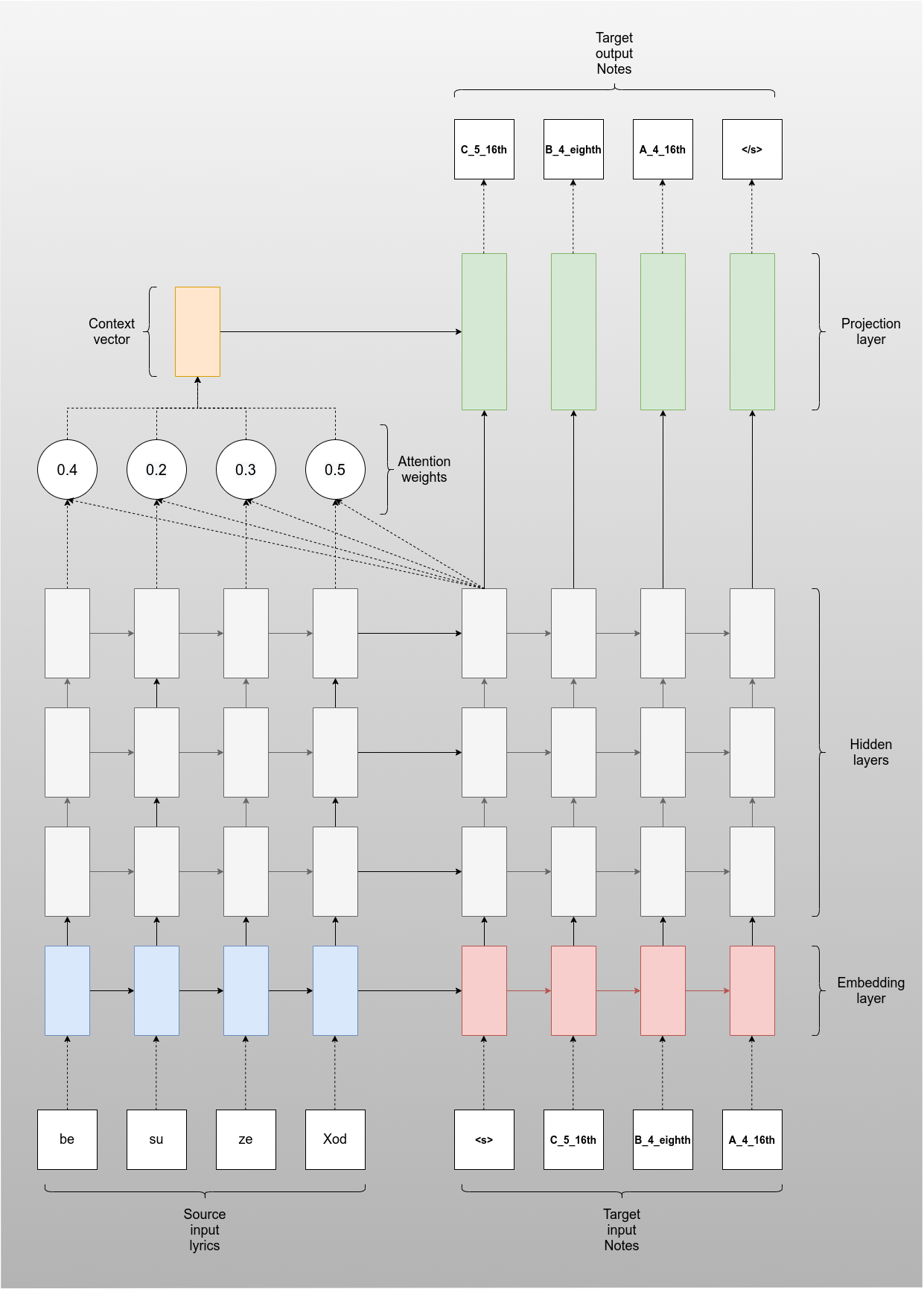}
\caption{Architecture of seq2seq model}\label{visina8}
\end{figure}

\par As shown in Figure 1, the architecture of our network consists following layers:

\begin{enumerate}
\item Embedding layer: ‌ It is the first row of the network in figure 1, where the sequential input values are converted to a vector with the size of vocab words.

\item Hidden layers: These layers consist of two parts. The first one is the  Encoding layer which is the four columns to the left of the network in figure 1, where the embedded values are converted into a vector in another space and prepared for decoding in the destination language (Music notes). The second one is the Decoding layer which the last hidden state of the previous layer is taken as input and decoded into a vector in the destination language.

\item Projection layer: Unlike the embedding layer, it visualizes the input vector equivalent to its words in the destination language.

\item Attention layer: In this layer, several tasks are done to remember more information about the previous training steps:

\begin{itemize}
\item Calculating the attention weights by comparing the latest hidden state of the encoding and decoding layer.
\item Calculating the context vector using the weighted average of the last state of the encoding layer based on the attention weights.
\item Calculating the attention vector by combining the context vector and the last hidden state of the decoding layer.
\item Transferring the attention vector for use in the next training step. 
\end{itemize}

\end{enumerate}

\section{Experiments}

\subsection{System Configuration}
In order to implement the proposed architecture, we used Tensorflow-GPU 1.15.0. Furthermore, we used a system with a 1.8ghz core i7 CPU, 16GB RAM, and a 2048MB Geforce MX250 as GPU to conduct the experiments.

\subsection{Dataset}
We consider lyric syllables as source language and note sequences as destination language. Almost every machine translation methods require a parallel corpus in two languages. A \emph{parallel} corpus is a dataset consisting of sentence pairs translated respectively in two languages. Table 1 consists of a few rows of the desired dataset.

\begin{table}[H]
\centering
\begin{tabular}{p{.2\textwidth} p{.2\textwidth}}
Lyric & Melody \\
\midrule
go le san gam go le san gam  & G-4-eighth A-4-eighth B-4-quarter A-4-half A-4-eighth G-4-eighth A-4-quarter G-4-half \\\hline
Ci be gam az de le tan gam  & G-4-eighth A-4-eighth B-4-quarter A-4-half A-4-eighth G-4-eighth A-4-quarter G-4-half \\\hline
me se Af tAb a ge bar man  & A-4-eighth B-4-eighth C-5-quarter B-4-half B-4-eighth A-4-eighth B-4-quarter A-4-half \\\hline
na tA bi sar da mo bi ran gam  & A-4-eighth G-4-eighth A-4-quarter G-4-quarter F-4-eighth G-4-eighth A-4-quarter B-4-quarter A-4-half \\\hline
\end{tabular}
\caption{Example of how the required parallel corpus looks like.}
\end{table}

Figure 2 indicates a note sheet from a song. The following challenges in digitally processing and storing a lyric's notes and syllables are considered.

\begin{figure}[H]
\centering
\includegraphics[width=.45\textwidth]{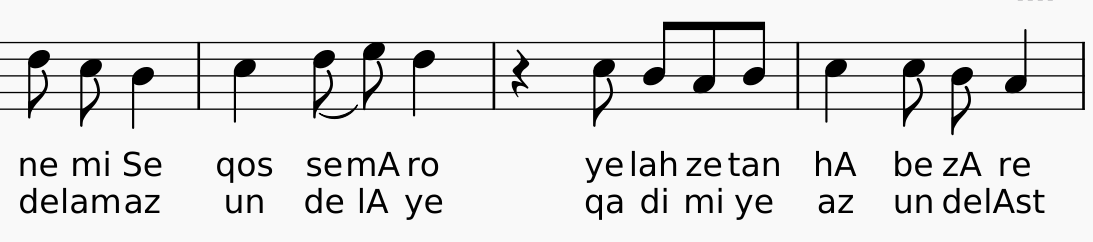}
\caption{A note sheet example}
\end{figure}

\subsubsection{The MusicXML Format to Store the Music Notes}

Digital sheets of music notes can be stored and transferred in different ways. The midi and MusicXML formats are the most widely used ones.

\par MIDI is a standard protocol in the music industry that connects digital musical instruments. MIDI files contain notes, sounds, measures, and other elements in a note sheet \cite{midi}.

\par MusicXML is an XML-based format to transfer and maintain the digital note sheets using standard XML tags so that various software can process these files \cite{musicxml01}.
\par The MusicXML format is selected due to the XML markup language's ease of processing and versatility.

\subsubsection{Collecting the Iranian Music Data}
To the best of our knowledge, any dataset on Iranian music pieces has not been collected and published digitally yet, and there are very few educational music books that collected some of these pieces. Considering our problem's scope, we approached the following solution: First, the printed pieces were scanned into PDF format, then the musical features get recognized using OMR software. Music Optical Reader (OMR) systems have the same function as OCRs. However, they recognize notes and other musical symbols from an image instead of natural language letters. This method requires human editing due to its limited accuracy, which takes less than an hour for each lyric.

This method requires a significant number of printed note sheets. Five music educational books consisting of famous pieces so that the lyric syllables were written separately for each note were used.

\par There are some well-known OMR systems like \cite{scanscore}, \cite{playscore} and Audiveris \cite{audiveris}. Audiveris is a great option because it is both open source and runs on all operating systems written in Java. This software can add OCR to the sign recognition process for extracting lyric syllables. There is a famous open-source OCR software, \cite{tesseract}, which we used in OMR.

\subsubsection{Human Editing}
The OMR system outputs have limited accuracy because of the differences in the letters used for lyrics and the quality of the scanned books. Hence, after converting each piece, it must be edited.

\par Note Sheets use the Latin alphabet to represent choral syllables. Some Persian letters lack the equivalent of a single letter in the Latin alphabet. We used a convention for annotating these letters.

\subsubsection{Sentence Definition in a Music Piece}
A parallel corpus is a collection of sentences and their translations. In natural languages, the sentence is well defined. However, a suitable definition of the sentence must be provided in music language.
There are various perspectives for defining the sentence in music, and the following four methods are considered:

\begin{itemize}

\item \textbf{Fixed length}: According to the successful parallel corpora published these years, sentences are on average five words, and their longest sentence does not exceed 35 words. Therefore, to define a sentence in music, every five consecutive syllables can be considered one sentence.
In this method, sentences are musically and linguistically incomplete. While this segmentation provided good data length for the model, it lacks semantic support.

\item \textbf{Between both Gaps in the lyric}: Another method is to consider the lyrics line as the basis of the decision and set the starting point of a sentence to the beginning of the lyric, and wherever there is no syllable accompanying a note, is the sentence end. This method guarantees the concept of the selected lyric piece as a sentence. However, in Iranian music, the \textbf{tremolo} technique is widely used. Tremolo is a wavering effect in a musical tone, produced either by a rapid reiteration of a note or by a rapid repeated slight variation in the pitch of a note. This technique is not displayed with a fixed standard in note sheets. In the process of digitizing pages, a syllable to be read in tremolo is a few notes away from the following syllable. As a result, a gap is created in the lyric, which is not the end of the sentence, making this method not applicable.

\item \textbf{Sequential measures}: Each rhythmic music piece consists of several measures containing several notes with equal total duration. A rhythmic piece cannot begin or end in the middle of a measure. From a musical point of view, each measure may be considered a sentence. In this way, it is possible to separate the musically meaningful parts; however, it does not assure that separated lyrics are meaningful.

\item \textbf{Silence in the melody}: According to a large number of dataset pieces, it was determined that wherever the melody reaches silence or the finish line, it separates a significant part of the sequence of notes. Meaningfulness here indicates that the beginning or end of a sequence of several notes is not unexpected or improper. In this method, the pieces are first separated, and each part is considered a \textbf{melodic sentence}. On the other hand, in combination with the first method (lyric gaps), the beginning of the lyric can be found in a melodic sentence. The sentence ends can be considered the end of the melodic sentence to avoid losing the tremolo notes. This code snippet \cite{jafari21} performs this operation.
\par For example, Figure 3 segments a sentence. It can be seen that each sentence is meaningful both in terms of lyric and melody.

\end{itemize}

\begin{figure}[H]
\centering
\includegraphics[width=.45\textwidth]{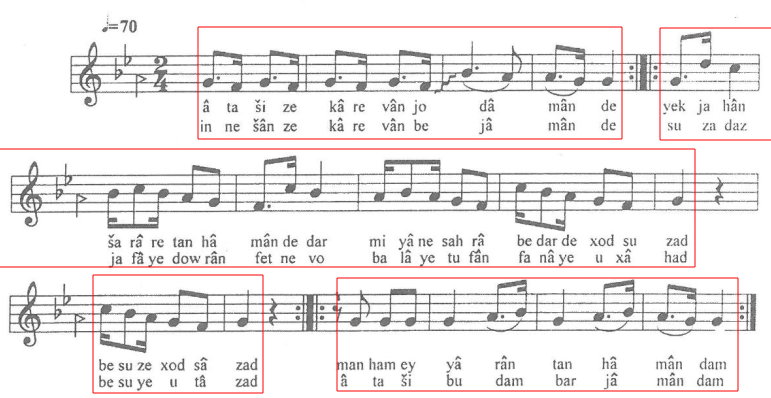}
\caption{An example of segmenting the musical sentences by silences in melody}
\end{figure}

\subsubsection{Preprocessing}

\par Feature selection: We know that a note contains numerous information and features. The following table presents all these features stored for each note as MusicXML.
There are many features provided with a note in MusicXML format, but just some of them can be useful in our case. The following features were extracted from the dataset: 
\begin{itemize}
\item \textbf{Scale}: Each Scale represents a scale of the diatonic signature, which is specified by letters A to G.
\item \textbf{Octave}: Specified by 10 degrees, where the middle degree specified the middle C scale.
\item \textbf{Type}: Note type is a positive fractional number that specifies the duration of a note as a fraction of a white note. i.e., 1/2, 1/4, 1/8.
\item \textbf{Text}: Single syllable of lyrics usually written in the phonetic alphabet.
\end{itemize}

\par As mentioned in the Methodology section, the rhythmic and melodic features along each other resolve the network requirements to learn the relationship between the notes and the syllables. The melodic features of a note are reflected in the pitch, comprised of Scale and Octave. The rhythmic features are also reflected in the note type or duration; for simplicity, note type is used here. On the other hand, the lyric is required to construct the parallel corpus.

\subsubsection{Statistics of Extracted Data}
Finally, 100 music pieces of five scanned books were compiled after character recognition and editing in MusicXML format. After pre-processing and separating the sentences, the required parallel corpus with the following specifications was used. The code snippet \cite{jafari21} implements the entire process of pre-processing MusicXML files. Some stats about the extracted data are described in Table 3.

\begin{table}[H]
\centering
\begin{tabular}{l r}
\midrule
Number of extracted sentences      & 1521  \\
The average syllables per sentence & 10.12 \\
Average notes per sentence         & 11.43 \\
Unique syllables                   & 775   \\
Unique notes                       & 105   \\
Vocabulary variety of syllables    & 0.05  \\
Vocabulary diversity of notes      & 0.006 \\
\bottomrule
\end{tabular}
\caption{Statistics of extracted data}
\end{table}

Extracted parallel corpus and 100 annotated MusicXML sheets of Iranian songs are publicly available at \cite{jafari2021-b}. We used 80\% of the data as training data, 10\% as development data, and the remaining 10\% for the test data.

\section{Evaluation}

The output should be examined from two points of view; first, the machine's viewpoint about its performance in detecting and repeating the patterns. Second, the qualitative viewpoint examines the pleasantness of machine-generated melodies for humans. These two viewpoints are described in the following.

\subsection{Machine Evaluation Using BLEU Score}
To evaluate the accuracy of a translator, a measure called BLEU is used. This measure is the abbreviation of "bilingual evaluation understudy" which is a method used for qualitative evaluation of a text translated by the machine \cite{papineni02}.

\par The translation quality is defined as the similarity of the machine output with human translation. This metric is one of the first metrics that suggest a high correlation with human judgment. Also, it is light in terms of processing. Thus, it is one of the widely applied translation evaluation methods.

\par To find the best hyperparameter, their impact should be evaluated under the same condition, and the more successful composition should be found among the most effective ones. 
Then five of the highest scores are selected, and the same models are trained for 10 epochs with 1000 steps.
\par Table 4 lists the shared hyperparameters among those five experiments, which also denotes that the model will produce lower BLEU scores by changing these values. We collected these values through grid search.

\begin{table}[H]
\centering
\begin{tabular}{@{}lr@{}}
\midrule
Number of units      & 128             \\
Direction            & One-directional \\
Using language model & No              \\
Inference method     & Greedy          \\
Cell type            & LSTM            \\
\bottomrule
\end{tabular}
\caption{Shared hyperparameters}
\end{table}

The Greedy inference method means choosing the nearest word to the decoded vector at each decoding step and feeding it to the next step.\cite{luong17}

\par Different hyperparameter settings for the five best BLEU score experiments are shown in Table 5.

\begin{table}[H]
\centering
\begin{tabular}{@{}cccc@{}}
\toprule
\textbf{\begin{tabular}[c]{@{}c@{}}Attention\\ type\end{tabular}} & \textbf{\begin{tabular}[c]{@{}c@{}}Attention\\ architecture\end{tabular}} & \textbf{\begin{tabular}[c]{@{}c@{}}Number\\ of layers\end{tabular}} & \textbf{\begin{tabular}[c]{@{}c@{}}Best \\ BLEU\end{tabular}} \\ \midrule
Bahdanau\cite{bahdanau14}                                                & GNMT                                                                         & 3                                                                      & 13.91                                                         \\
Bahdanau                                                          & Standard                                                                     & 4                                                                      & 15.31                                                         \\
Bahdanau                                                          & GNMTv2                                                                       & 4                                                                      & 14.47                                                         \\
Bahdanau                                                          & GNMT                                                                         & 4                                                                      & 11.62                                                         \\
No attention                                                      & Standard                                                                     & 4                                                                      & 16.47                                                         \\ \bottomrule
\end{tabular}
\caption{Five best BLEU score experiments}
\end{table}

\par As a result of this experiment, three more successful models are used to generate human evaluation pieces. In the future, to validate this method, probabilistic models like CRF can also be trained on this data and compare the BLEU score with the neural network.

\subsection{Human Evaluation}
Evaluating an artwork is difficult and inaccurate. Thus, taste and cultural characteristics significantly affect perception from a work of art. The output of a generative
music model cannot be evaluated solely by quantitative measures. Because these measures are based on a hypothesis that suggests there is a significant relationship between the lyrics (or natural language) and the melody.
\par Reviewing similar articles related to this topic, it is observed that various methods have been presented for human evaluation. One of the most valuable methods for this topic is a questionnaire based on the Likert Scale. Likert Scale is used to measure emotions, ideas, satisfaction, and familiarity, in Psychometrics. Likert includes different score scales with 5 or 7 options for bipolar concepts (like satisfaction or dissatisfaction). The last option is the agreement with the other end of the pole (maximum satisfaction). Other options are within this range \cite{krosnick12}. This method has been used to investigate the pleasantness of the listener to a machine output or a h	uman piece. Also, for the problem of generating melody for a lyric, this method can be used to measure the fitness of the output with the input.

\par The model's output is a piece of melody for one row lyric as the input. Therefore, two aspects should be evaluated. First is the quality of the generated melody; second, the fitness of the melody with the input lyrics. The five-point Likert Scale is used To measure these two.
\par Furthermore, other issues require investigation, like the familiarity of the questionnaire audience with music or previous familiarity with the samples. These two aspects are also included in the questionnaire as two-option questions.
\par Finally, the questionnaire is designed for the machine and human-generated pieces with the same lyrics. The questions are asked independently for each piece:
\begin{itemize}
\item "Are you familiar with solfege?"
\item "Have you ever heard a music with this lyric?"
\item "If you are asked to score the melody of this song, apart from other components like the singer's voice or the instruments, what is your score from 1 to 5?"
\item "How much does the melody of this piece fit its lyrics? (for example, a happy melody is not suitable for a sad lyric). Score from 1 to 5."
\end{itemize}

\par Fourteen samples are selected for evaluation. Each sample is a piece of lyrics about four lines long. These lyrics are selected from "Classic Literature-Classic Music" and "Contemporary Literature-Pop Music" classes. Also, it is tried to select pieces of music that are not very well-known so that the previous familiarity does not bias the audience. On the other hand, they are selected from outstanding lyricists and singers to ensure a minimum quality of the lyrics and the song. The lyrics are given as input to the model, and the suggested melodies by the system are saved in MusicXML format publicly accessible at \cite{jafari2021-a}.

\par The system output is performed and recorded with a professional singer. While performing the pieces, a maximum change of 10\% was applied to the melody by the singer. We only used the duration of the notes as a rhythmic feature in the training process. Therefore, the outputs lack the time signature (specifying the rhythm number of the piece). Thus, the time signature of the pieces is selected by the singer.

\par To validate the ideas, it is required to consider the ground truth. To this end, the original songs are also surveyed. Piano and human vocals are used to minimize the difference between the conditions of the original piece and the generated one. Conditions refer to considering different components of a complete piece of music, for instance, the employed instruments, background chords, and studio editions. 
Original and recorded pieces are also publicly accessible in \cite{jafari2021-a}.
	
\subsection{Results Analysis}
Fourteen generated pieces are shared among different audiences in 4 different questionnaires. Totally 174 evaluations are collected, and an average of 30 individuals have evaluated each piece. In the following, the results of the questionnaires are analyzed.
\par As shown in Figure 4, the system outputs have obtained an average score of 3 from 5, and the original pieces have obtained 4. This result might be promising for taking the next steps towards solving this problem.

\begin{figure}[H]
\centering
\includegraphics[width=.4\textwidth]{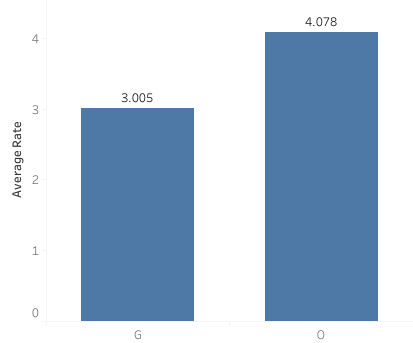}
\caption{The average score given by the audience to the pleasantness of the generated melody (G) and the human-made melody (O) from one to five.}
\end{figure}

\par As shown in Figure 5 and the pleasantness score for each generated piece, it is seen that piece b3 (Nameye Jodayee) has obtained the highest score and minimum distance with the original piece. Also, piece b2 (Mey-E-Eshgh) has achieved the maximum distance with the original piece, and piece b1 (Masiha) has obtained the minimum score among other pieces.

\begin{figure}[H]
\centering
\includegraphics[width=.4\textwidth]{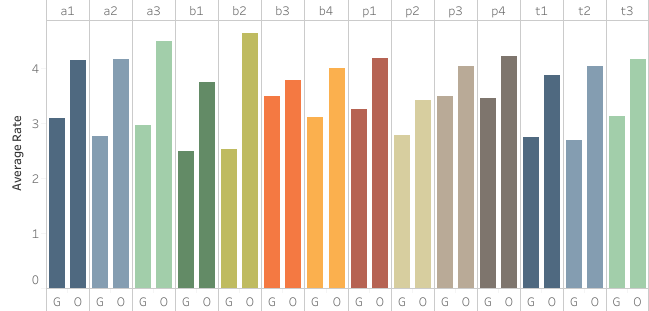}
\caption{The average score of the pleasantness of each piece}
\end{figure}

By comparing the average score of pleasantness and fitness of the human-made pieces in Figure 6, it can be seen that the classic style is more popular than the pop style, but it is contrary to machine-made pieces. Here, it can be concluded that the music style of the training data affects the output significantly because most training data was of pop style. Therefore, an essential step in future studies would be separating training data style. Also, it can be predicted that constructing classical music using a machine is more complicated, which might be due to different reasons, like using a quarter-tone that is common in Iranian classical music. However, quarter-tone is not defined in available software and infrastructure.

\begin{figure}[H]
\centering
\includegraphics[width=.4\textwidth]{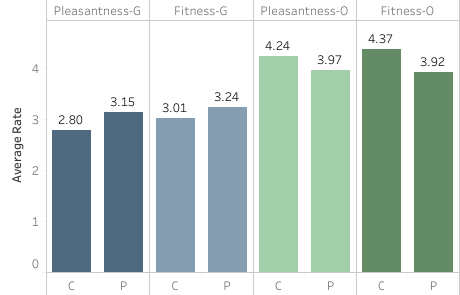}
\caption{The average pleasantness and fitness per music style. (C: classic music, P: pop music, Pleasantness-G and Fitness-G represent pleasantness and fitness of the generated pieces, and Pleasantness-O and Fitness-O indicate pleasantness and fitness of the original pieces made by humans.)}
\end{figure}

The first question of the questionnaire asks if the audience is familiar with solfege. It can be said that familiarity with solfege is highly related to the audience's familiarity with the music. Solfege is necessary to play an instrument. The one who plays an instrument is familiar with the melody, rhythm, and other music components. Therefore, familiarity with solfege is the minimum familiarity with the music. It can be expected from the audience familiar with the music to present a more accurate comment about the pieces they hear. In this survey, about \textbf{40\%} of the audience had the minimum familiarity with the music.

\par In Figure 7, the audience's points of view regarding pleasantness and fitness of the pieces are compared for the audience familiar with the music and those unfamiliar with the music. The ideas are almost similar for the human-made pieces with a high score, ensuring that the selected pieces are of high quality. However, for the machine-made pieces, there is a significant difference between the ideas of these two groups. The pleasantness of the machine-made pieces is scored lower by the audience familiar with the music.

\begin{figure}[H]
\centering
\includegraphics[width=.4\textwidth]{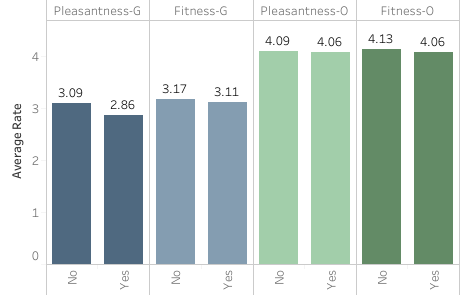}
\caption{The average score of pleasantness and fitness for the audience familiar and unfamiliar with solfege}
\end{figure}

\section{Conclusion}
\subsection{Key Results}

Acceptable results are obtained in three different classes considering the objectives of this study:
\begin{itemize}
\item Digital Iranian music data: In this thesis, we could collect more than 100 pieces of Iranian music in different styles and store them digitally in standard MusicXML format. We published the dataset to help the interdisciplinary studies of computer science and music continue.

\item Presenting a method based on translation for generating melody:
In this research, inspired by \cite{bao19}, a deep seq2seq neural network for generating melody based on the given lyrics was developed. The outcome is a system that can be trained on music data of any language and create a melody for its lyrics. It can be said that this system is the first automatic melody composer for Persian lyrics by collecting data from Iranian music.

\item Human evaluation:
In this paper, the digital output of the machine is performed and recorded as an audio file provided to the audience for human evaluation. The outcome is more than 20 minutes of music in 14 pieces sung by a professional vocalist in the studio.
\end{itemize}

\subsection{Future Works}

\subsubsection{Defining Persian Classic Music Dastgah}
A unique feature of Persian Classical Music is its "Dastgah"s. Each Dastgah has different rules about using semitones and quarter-tones, which can be used to detect the Dastgah from training data. Apart from these laws, each Dastgah includes a set of "Gusheh"s with specific semantic contents. Each Dastgah also describes emotional contents. For example, one Dastgah might describe heroic emotions, and the other might describe sorrow and grief.
\par The first challenge in this context is using quarter-tone in Persian Classical Music dastgah. Because using quarter-tone is not common in Western music and the musical symbol "Koron" that represents quarter tone cannot be easily defined in MusicXML format.
\par Besides, a natural language classification system can be trained for each Dastgah separately by the lyrics of these pieces. This subsystem can suggest the proper Dastgah for the input lyric.

\subsubsection{Automatic Suggestion of Time Signature}
The time signature is a musical notation that specifies how beats are contained in one measure. In other words, the essential element for rhythm suggestion is selecting the time signature and configuration of the notes in the times. On the other hand, rhythm is an essential element in creating emotions in the music audience. Therefore, by emotional classification of the lyric along with the rhythmic classification of the training data, the possibility of specifying the time signature and the rhythm of the constructed piece is provided for the system.

\subsubsection{Automatic Scale Suggestion for Western Music and Pop Style}
Scale in music is the set of notes sorted based on their pitch. The scales are divided into two main groups of major and minor, where the first one creates happiness and the second one creates sadness in the audience. 
For Scale suggestion, again by semantic analysis of the lyrics, major or minor Scale can be determined first. Then, the initial octave note can be suggested based on the training data.

\bibliographystyle{plain}
\bibliography{iccc}

\end{document}